\documentclass[doublecol]{epl2} 
\usepackage{amsfonts}

\newcommand{\bv}[1]{{\boldsymbol #1}}

\title{Lane formation in a lattice model for oppositely driven binary particles}
\shorttitle{Lane formation in a lattice model for oppositely driven binary particles} 

\author{H. Ohta}
\shortauthor{H. Ohta}

\institute{LPTMS, CNRS and Universit\'e Paris-Sud, 91405 Orsay Cedex, France}

\abstract{
Oppositely driven binary particles with repulsive interactions 
on the square lattice are investigated at the zero-temperature limit. 
Two classes of steady states related to stuck configurations and lane formations
have been constructed in systematic ways under certain conditions. 
A mean-field type analysis carried out using a percolation problem based on the constructed steady states 
provides an estimation of the phase diagram, which is qualitatively consistent with numerical simulations.
Further, finite size effects in terms of lane formations are discussed.}
\begin{document}

\maketitle

\section{Introduction}
Nonequilibrium systems exhibit collective phenomena under certain conditions.
The understanding of the relationship between such collective phenomena and 
microscopic properties of the system 
has been an important topic in statistical physics.
Recently, it has been observed that 
various systems consisting of oppositely driven binary particles with repulsive interactions exhibit a collective phenomenon called lane formation, 
where the same type of particles align to the driven direction 
\cite{Lowen1,Netz,Lowen11,Lowen2,Lek,Lowen3,Lowen4}. 
The observation of lane formations in various systems 
suggests that certain types of properties are universal in lane formations, irrespective of the systems.

Thus far, a phenomenological theory for determining a special condition 
to show lane formations in colloidal suspensions has been presented, which is consistent to 
numerical results in some parameter regions \cite{Lowen11,Lowen2}. 
However, we still lack the knowledge for the relationship 
between lane formations and microscopic properties.
For examples, questions on whether the special condition is related 
to some nonequilibrium phase transitions 
or which types of fluctuations, for instance, finite size effects, 
appear in lane formations remain unanswered.

As is common in the literature on statistical physics, 
lattice models are often useful for obtaining insights into universal phenomena, 
owing to the simplicity of the system and the universality itself.
For example, driven lattice gas models and simple exclusion processes 
have provided many insights into universal aspects of nonequilibrium systems,
such as phase transitions and the KPZ universality \cite{Spohn,Derrida,Sasamoto}. 
In this context, a simple lattice model exhibiting a lane formation can be 
useful for understanding lane formations more comprehensively. 
However, extensive studies have not been conducted on this topic.

In this paper, in order to obtain insights into the universal aspects of lane formations, 
we propose a lattice model where binary particles with purely repulsive interactions
are driven in opposite directions.
We present the exact constructions of two classes of steady states 
related to stuck configurations and lane formations.
On the basis of these constructions, 
we introduce a percolation problem in a stochastic cellular automaton, 
which helps us estimate the phase diagram. 
Further, we discuss finite size effects in lane formation using the constructed steady states.

\section{Model}
\begin{figure}
\centering
\onefigure[width=6.0cm,clip]{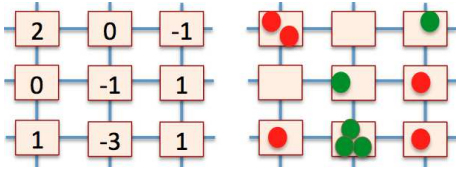}
\caption{Schematic of the model showing
the correspondence between the configuration of $\{\sigma_i\}_i$ (left) and that of binary particles (right).}
\label{rule0}
\end{figure}

\begin{figure}
\centering
\includegraphics[width=2.8cm,clip]{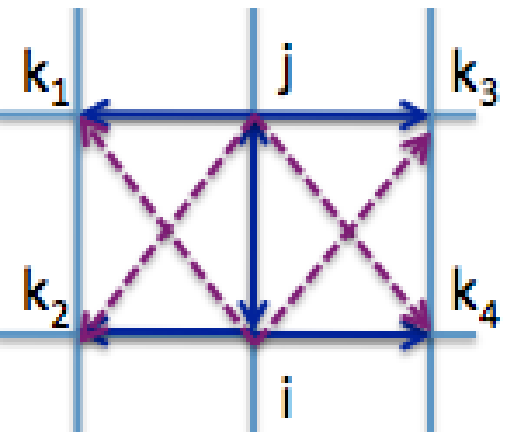}
\includegraphics[width=2.3cm,clip]{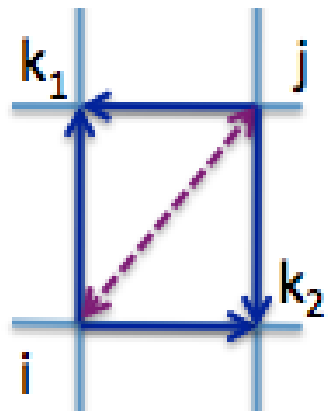}
\caption{Hopping from $i$ to $j$ with $l_{\rm hop}=\sqrt{2}$: 
$\Delta E_{ij}^0(\bv{\sigma})$ is calculated by $\Delta E_{ij}(\bv{\sigma})$,$\Delta E_{ik_1}(\bv{\sigma})$, 
$\Delta E_{ik_2}(\bv{\sigma})$, $\Delta E_{ik_3}(\bv{\sigma})$, $\Delta E_{ik_4}(\bv{\sigma})$, 
$\Delta E_{ji}(F_{ij}\bv{\sigma})$, $\Delta E_{jk_1}(F_{ij}\bv{\sigma})$, 
$\Delta E_{jk_2}(F_{ij}\bv{\sigma})$, $\Delta E_{jk_3}(F_{ij}\bv{\sigma})$, 
and $\Delta E_{jk_4}(F_{ij}\bv{\sigma})$ (left), 
or $\Delta E_{ij}(\bv{\sigma})$, $\Delta E_{ik_1}(\bv{\sigma})$, $\Delta E_{ik_2}(\bv{\sigma})$,
$\Delta E_{ji}(F_{ij}\bv{\sigma})$, $\Delta E_{jk_1}(F_{ij}\bv{\sigma})$, 
and $\Delta E_{jk_2}(F_{ij}\bv{\sigma})$ (right).  }
\label{rule1}
\end{figure}

Let us consider a square lattice $\Lambda$ consisting of site 
$i\in \{\{i_x,i_y\}\in \mathbb{N}\times\mathbb{N} | 1\le i_x,i_y\le L\}$ 
where $\mathbb{N}$ is the set of natural numbers. 
In order to introduce binary (positive and negative) particles on the lattice, 
we consider an occupation variable $\sigma_i\in \mathbb{Z}$ $(-L^2\le\sigma_i\le L^2)$, 
where $\mathbb{Z}$ is the set of integers.
In this expression, $\sigma_i$ indicates that there are $|\sigma_i|$ number of particles 
with the sign of $\sigma_i$ at site $i$, and no particles if $\sigma_i=0$ 
as illustrated in figure \ref{rule0}.
Therefore, the density $\rho$ of the particles in the system is 
defined as $\rho\equiv\frac{1}{L^2}\sum_{i\in\Lambda}|\sigma_i|$.
We consider a situation where each particle interacts 
with soft-core repulsive interactions with the following Hamiltonian 
for ${\bv \sigma}\equiv\{\sigma_i\}_{i\in\Lambda}$:
\begin{eqnarray}
H(\bv{\sigma})=V_0\sum_{i\in\Lambda}|\sigma_i|(|\sigma_i|-1),\label{ham}
\end{eqnarray} where $V_0>0$. On the basis of this purely repulsive Hamiltonian, 
we consider continuous-time Markov processes
where a particle hops to another site with some transition rates; 
this process is expressed as follows. 
Preliminarily, in order to express the hopping of one particle from site $i$ to site $j$, 
we define an operator $F_{ij}$ that satisfies $F_{ij}\sigma_i=\sigma_i-{\rm sgn}(\sigma_i)$, 
$F_{ij}\sigma_j=\sigma_j+{\rm sgn}(\sigma_i)$, and $F_{ij}\sigma_k=\sigma_k$, 
where ${\rm sgn}(x) = - 1$ for $x<0$ and ${\rm sgn}(x) = 1$ for $x>0$, otherwise ${\rm sgn}(x) = 0$.
Here, the destination site $j$ can be in set ${\rm B}_i\equiv \{ j\in \Lambda | |i-j| \le l_{\rm hop} \}$ 
where $|i-j|\equiv\sqrt{(i_x-j_x)^2+(i_y-j_y)^2}$ and $l_{\rm hop}\ge 1$, as shown in figure \ref{rule1}.
The transition rate $R(\bv{\sigma} \to F_{ij}\bv{\sigma})$ 
between state $\bv{\sigma}$ and $F_{ij}\bv{\sigma}$ with $\sigma_i\neq 0$ is
\begin{eqnarray}
\Theta(\sigma_i\sigma_j)w_{ij}(\sigma_i,\sigma_j)r(\bv{\sigma}\to F_{ij}\bv{\sigma}), \label{rate}
\end{eqnarray} 
with $r(\bv{\sigma}\to F_{ij}\bv{\sigma})\equiv$
\begin{eqnarray}
& \exp(-\beta/2(\Delta H_{ij}(\bv{\sigma}) - D_{ij}(\sigma_i) 
- \Delta E_{ij}^{0}(\bv{\sigma}))), \label{rate} \\
&w_{ij}(\sigma_i,\sigma_j)\equiv d_{ij}\frac{|\sigma_i+\sigma_j|}{2},
\end{eqnarray}
where 
 $\Delta H_{ij}(\bv{\sigma}) \equiv (H(F_{ij}\bv{\sigma})-H(\bv{\sigma}))/|i-j|
$, $D_{ij}(\sigma_i) \equiv f (j_x-i_x){\rm sgn}(\sigma_i)/|i-j|$, 
and $\Theta$ is a step function such that $\Theta(x)=1$ 
for $x\ge 0$, otherwise $\Theta(x)=0$.
We define $r(\bv{\sigma} \to F_{ij}\bv{\sigma})=0$ with $\sigma_i= 0$ 
to maintain consistency in this description. 
Further, we assume that $\Delta E_{ij}^0(\bv{\sigma})=\Delta E_{ji}^0(F_{ij}\bv{\sigma})$ 
holds; an explicit form of this expression is provided later.
According to these properties, 
one can prove that $R(\bv{\sigma} \to F_{ij}\bv{\sigma})$ 
satisfies the detailed balance condition with $f=0$. 


Here, we explain the physical meaning of each term. 
First, the term $D_{ij}(\sigma_i)$ governs the tendency of positive (negative) 
to hop in the positive (negative) direction along the $x$-axis 
with a driving field $f$. 
Second, the weight $w_{ij}(\sigma_i,\sigma_j)$ along the edge for sites $i$ and $j$ 
has two physically reasonable implications. 
One implication is to assume that the diffusion of a free particle 
is uniform for the $x(y)$ and diagonal directions.
Explicitly, we can realize this situation by considering 
that $|i-j|d_{ij}$ is independent of $j$, for example, using $\sum_{j\in{\rm B}_i}|i-j|d_{ij}=4$, 
which leads to $d_{ij}=1$ with $l_{\rm hop}=1$ or
$d_{ij}=|i-j|^{-1}/2$ with $l_{\rm hop}=\sqrt{2}$. 
It should be noted that $d_{ij}=d_{ji}$ by definition.
The other implication is that the term $|\sigma_i+\sigma_j|$ ensures that 
the number of hopping events at the edge for sites $i$ and $j$ occur 
in proportion with the number of particles along the edge. 
Third, the step function $\Theta(\sigma_i\sigma_j)$
ensures that every type of particle is conserved.

Here, we consider an explicit form of $\Delta E_{ij}^0(\bv{\sigma})$.
In order to construct a model with universal properties, 
we develop a guiding principle for determining forms of $\Delta E_{ij}^0(\bv{\sigma})$ 
in the following manner.
The idea is to consider the systems described by overdamped Langevin equations
where particles move in the exact direction of the driving force at the infinite driving force limit 
\cite{Lowen1}. Keeping this in mind, 
we assume that particles at each site $i$ hop only to sites 
$k\in {\rm B}_i$ with maximum $D_{ik}(\sigma_i)\Theta(\sigma_i\sigma_k)$
at the infinite driving field limit, which we refer to as {\it normality of the limit}.
For example, the following expression provides the {\it normality of the limit};
\begin{eqnarray}
\Delta E_{ij}^0(\bv{\sigma}) \equiv \min_{k\in {\rm B}_{ij}}
\{\Delta E_{ik}(\bv{\sigma}), 
\Delta E_{jk}(F_{ij}\bv{\sigma})\},
\end{eqnarray} 
where $\Delta E_{ik}(\bv{\sigma})\equiv \Delta H_{ik}(\bv{\sigma}) - D_{ik}(\sigma_i)$, and 
${\rm B}_{ij}\equiv\{k \in \Lambda| |k-i|, |k-j|\le l_{\rm hop}\}$ with $l_{\rm hop}=\sqrt{2}$. 
This expression also satisfies
$\Delta E_{ij}^0(\bv{\sigma})=\Delta E_{ji}^0(F_{ij}\bv{\sigma})\le 0$ 
and $0\le r(\bv{\sigma}\to F_{ij}\bv{\sigma}) \le 1$. In this study, we investigate only this case.
In order to capture a physical image of $\Delta E_{ij}^0(\bv{\sigma})$, 
we show $\Delta E_{ik}$ required to calculate $\Delta E_{ij}^0(\bv{\sigma})$ in figure \ref{rule1}.
It should be noted that in order to ensure that 
only $\Delta E_{ij}^0(\bv{\sigma})=\Delta E_{ji}^0(F_{ij}\bv{\sigma})$ holds, 
we can set $\Delta E_{ij}^0(\bv{\sigma})$ as a constant.
However, in this case, the model does not follow the {\it normality of the limit}.
As a result, the arguments mentioned in the rest of this paper do not hold. 
On the other hand, we expect that the arguments mentioned in the rest of this paper are robust 
against small changes in the expression for $\Delta E_{ij}^0(\bv{\sigma})$ if the expression 
provides the {\it normality of the limit}; however the extent of changes for which the argument hold 
is uncertain.
This type of guiding principle used to determine transition rates with universal properties 
has been discussed in different nonequilibrium lattice models with a nonconserved variable \cite{Hucht}. 

In this study, we consider the behaviours of the system 
by changing the set of parameters $(\rho,f)$.
Further, for simplicity, we focus on the case with the zero-temperature limit, 
the periodic boundary condition, and $50$:$50$ binary mixtures where 
the densities of positive and negative particles are $\rho/2$.
If $f=0$, the steady states are determined 
by the canonical distribution of Hamiltonian (\ref{ham}).
In particular, one can immediately determine that if $f=0$ and $\rho\le 1$, 
$\sigma_i$ in equilibrium states randomly takes only the values of $1$, $-1$, or $0$  
 with a fixed density $\rho$.
Obviously, there are no singular points in the parameter space under this equilibrium condition. 
We consider relaxation behaviours from this equilibrium initial condition.
It should be noted that these behaviours are 
still non-trivial even under these simple conditions owing to purely nonequilibrium effects, 
as explained later.
While performing Monte Carlo simulations for this model, we randomly select a particle. 
If it is located at site $i$, we select any $j\in {\rm B}_i$ with probability $d_{ij}/(2+\sqrt{2})$. 
Then, both the directions from $i$ to $j$ and from $j$ to $i$ are adopted with a probability of $1/2$.
Finally, a particle hops to the adopted direction (if $j \to i$) 
with the probability $\Theta(\sigma_i\sigma_j)r(\bv{\sigma}\to F_{ji}\bv{\sigma})$, 
and if there are no particles at site $j$, nothing occurs. 
This process is repeated and time $t=1$ corresponds to the repeated $L^2\rho/\tau_0$ steps 
where $\tau_0=1/(4+2\sqrt{2})$.

\section{Exact arguments and the related conjectures}\label{rig}
Let us confirm some non-trivial arguments for the model. 
Preliminarily, we consider a current-like quantity as follows. 
Let us define a colored current $J(t)$ at time $t$ with $\bv{\sigma}$ as
\begin{eqnarray}
J(t)\equiv\frac{1}{L^2\rho}\sum_{i\in\Lambda}\sum_{j_x=i_x\pm 1}
\Theta(\sigma_i\sigma_j){\rm sgn}(\sigma_i)r(\bv{\sigma} \to F_{ij}\bv{\sigma}),
\end{eqnarray} where $-1\le J(t)\le 1$ by definition. 
We consider $\mathcal{J} \equiv J(\tau)$ such that 
$J(\tau)=\lim_{\tau_{\rm o}\to\infty}\frac{1}{\tau_{\rm o}}\sum_{t'=\tau}^{\tau+\tau_{\rm o}}J(t')$. 
It is trivial that if $f=0$, 
$\mathcal J$ should be absolutely zero with  $\tau\to\infty$ owing to the symmetry of the directions.
By using this quantity, we define the maximum-current ({\it MC}) steady states as ${\mathcal J}=1$ 
and the stuck steady states as ${\mathcal J}=0$. 
Next, using the colored current, 
we construct the microscopic characterizations of such steady states in some parameter regions. 
After these constructions, we present some conjectures related to the constructed steady states.

\begin{figure}
\centering
\includegraphics[width=4.8cm,clip]{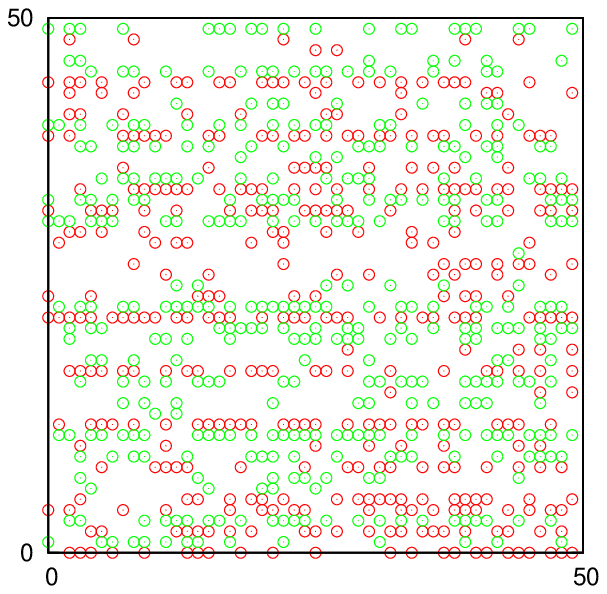}
\caption{Particle configuration of a numerically observed global-{\it MC} steady state at 
$f=100.0$ and $\rho=0.5$ with $L=50$. The red (green) circles are positive (negative) values 
for the occupation variable.}
\label{image1}
\end{figure}

\begin{figure}
\centering
\includegraphics[width=4.8cm,clip]{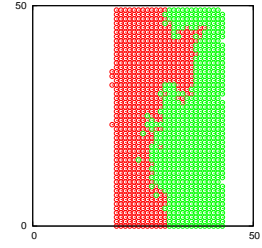}
\caption{Particle configuration in a numerically observed stuck steady state at $f=1.0$ and $\rho=0.5$ with $L=50$.}
\label{image2}
\end{figure}

\subsection{One explicit class of stuck steady states with $0<f<2V_0$ for $\rho <1$}
Let us construct one explicit class of the stuck steady states at $0<f<2V_0$ in the following manner.
We consider an auxiliary variable $\hat{\sigma}_i\in\{-1,0,1,2\}$ as follows.
\begin{eqnarray}
&\{\hat{\sigma}_i=2\}\equiv \{\sigma_{i_x,i_y}=1,\sigma_{i_{x}+1,i_y}=-1\},\\
&\{|\hat{\sigma}_i|\le 1\}\equiv \{\sigma_i=\hat{\sigma_i}\}.
\end{eqnarray}
On the basis of this variable $\{\hat{\sigma}_i\}_i$, we consider a cellular automaton 
for determining the value of $\hat\sigma_i$ under the initial condition $\hat{\sigma}_{i^{\rm o}}=2$ 
and $\hat{\sigma}_{i}=0$ for all the other sites $i$; 
the rule to update the values for all the other sites is explained in the following sentences. 
First, we set $\hat{\sigma}_{i_x^{\rm o}-1,i_y^{\rm o}+1}=2$ or $\hat{\sigma}_{i_x^{\rm o},i_y+1}=2$ 
or $\hat{\sigma}_{i_x^{\rm o}+1,i_y^{\rm o}+1}=2$.
Next, if $\hat{\sigma}_{i_x,i_y^{\rm o}+1}=2$, we set 
$\hat{\sigma}_{i_x-1,i_y^{\rm o}+2}=2$ or $\hat{\sigma}_{i_x,i_y+2}=2$ 
or $\hat{\sigma}_{i_x+1,i_y^{\rm o}+2}=2$.
This process is repeated. After setting $\sigma_i=2$ for at least one site 
for every values of $i_y$, we set $\hat{\sigma_i}=-1$ for $i_y^{\rm l} <i_x\le i^{\rm r}$, 
where $i_y^{\rm l} \equiv \max_x i_x + 1$ such that $\hat{\sigma}_i=2$ for a given value of $i_y$ 
and $i^{\rm r}\equiv\max_y i_y^{\rm l} + 1$.
Similarly, we set $\hat{\sigma_i}=1$ for $i^{\rm l}\le i_x<i_y^{\rm r}$, where $i_y^{\rm r} \equiv \max_x i_x$ 
such that $\hat{\sigma}_i=2$ for a given value of $i_y$ and $i^{\rm l}\equiv\max_y i_y^{\rm r}$. 
Thus, we obtain a sequence $\{\hat{\sigma}_i\}_i$.
Finally, if we set $H(\bv{\sigma})=0$, 
this system reaches a steady state with ${\mathcal J}=0$ 
at the long time limit because $\{\hat{\sigma}_i\}_i$ are 
independent of $t$ in this construction. 
This corresponds to the fact that there exist the stuck steady states
characterized by $\{\hat{\sigma}_i\}_i$ with $H(\bv{\sigma})=0$ for 
$f<2V_0$ and $\rho=\frac{1}{N}\sum_i|\sigma_i|$.
Further, a state constructed by using $\{\hat{\sigma}_i\}_i$ and $\{\hat{\sigma}_i'\}_i$ 
with $\hat{\sigma}_i'\hat\sigma_i=0$ and $H(\bv{\sigma})=0$ 
can also be a stuck steady state. 
We define the stuck steady states constructed in this way as 
{\it CA}-stuck steady states. However, there are no principles 
to determine whether {\it CA}-stuck steady states appear as steady states 
under the equilibrium initial condition, which is discussed later.

\subsection{One explicit class of maximum-current steady states
with sufficiently large values of $f$ for any density}
Let us construct one explicit class of the maximum-current steady states
with sufficiently large values of $f$ for any density, as follows. 
First, let us define 
$N_{{\rm p},y}\equiv \sum_{i_x'}|\sigma_i|\Theta(\sigma_{i'})\delta_{i_x',y}$ 
and $N_{{\rm n},y}\equiv \sum_{i_x'}|\sigma_{i'}|\Theta(-\sigma_{i'})\delta_{i_x',y}$.
For each value of $y$, we set $N_{{\rm p},y}$ and $N_{{\rm n},y}$ such that $N_{{\rm p},y}N_{{\rm n},y}=0$ 
under the condition $\sum_yN_{{\rm p},y}=\sum_yN_{{\rm n},y}=\rho L^2/2$.
Then, if we set $f$ such that 
\begin{eqnarray}
f \ge \frac{2\sqrt{2}V_0}{\sqrt{2}-1}
\max_y(N_{{\rm p},y}, N_{{\rm n},y}), \label{ff}
\end{eqnarray} we can easily find that this system reaches a steady state ${\mathcal J}=1$ 
at the long time limit because $N_{{\rm p},y}$ and $N_{{\rm n},y}$ are independent of $t$.
This corresponds to the fact that 
the set $\{N_{{\rm p},y},N_{{\rm n},y}\}_{0\le y\le L}$ under the condition mentioned above 
determines a maximum-current steady state for $f$ satisfying (\ref{ff}) and 
$\rho=\frac{1}{N}\sum_{y=0}^L(N_{{\rm p},y}+N_{{\rm n},y})$. 
We define the maximum-current steady state constructed in this way as 
global-{\it MC} steady states. However, there are no principles to 
determine whether global-{\it MC} steady states appear as steady states 
under the equilibrium condition, which is also discussed later.

\subsection{Conjecture A: Mean-field type analysis of phase transitions}\label{con}
So far, we have  microscopically  constructed
the {\it CA}-stuck steady states and the global-{\it MC} steady states, 
which lead us to expect that singular points of $\mathcal{J}$ exist in parameter space $(\rho,f)$
where the steady state changes qualitatively. 
With this background, let us assume that on fixing $\rho$,
there is a line $f_{\rm c}(\rho)$ where $\mathcal{J}$ is singular in driving field $f$. 
On the basis of this assumption, we attempt to approximately estimate $f_{\rm c}(\rho)$ as follows.

In low-density regions, particles appear to be driven independently.
Therefore, a non-zero value of $f_{\rm c}(\rho)$ would be observed only in sufficiently 
high-density regions, and there should be a threshold density $\rho_{\rm th}$ 
below which $f_{\rm c} = 0$. 
We estimate $\rho_{\rm th}$ as follows. 
First, it should be noted that it is plausible for a type of seed 
to exist under the initial condition for realizing the {\it CA}-stuck steady states. 
One reasonable candidate for such a seed is the particles present at the sites 
indicated by the one-step run of the cellular automaton from a focused particle, 
which we refer to as an {\it automaton-pointed segment} of the focused particle, 
as discussed in the construction of the {\it CA}-stuck steady states.
We assume that if there exists at least one sequence of 
automaton-pointed segments starting from a focused particle, 
which connects with a site with distance $L$ from the focused particle under the initial condition, 
the system can reach a {\it CA}-stuck steady states after the dynamical particle-exchange processes.
As mentioned in the construction of the {\it CA}-stuck steady states, this 
consideration is applied in the case of $f<2$, which leads to $f_{\rm c}(\rho_{\rm th})=2V_0$.
On this assumption, we consider the probability $P_{y}$ that there exists
at least one sequence of automaton-pointed segments starting from a focused particle at $i_y=y$, 
which connects with a site with $i_y= L$ under the initial condition. 
This is a typical percolation problem in a stochastic cellular automaton 
with a percolation point $\rho_{\rm p}$ \cite{Hin}. 
Reminding that the number of sites in such an automaton-pointed segment is $4$, 
if each site in an automaton-pointed segment assumed to be independent, 
we can obtain the recursive equation $P_{y}=\rho(1-(1-P_{y+1})^4)$. 
Thus, we can estimate a lower bound $\rho_{\rm p}^{\rm l}=1/4$ of 
the percolation point $\rho_{\rm p}$, which also gives an approximate value of $\rho_{\rm th}$. 
On the basis of this consideration, let us estimate $f_{\rm c}(\rho)$ at $\rho>\rho_{\rm th}$. 
Here, we consider the condition with which the percolation breaks owing to the hopping of a particle 
through an automaton-pointed segment, where only the same type of particles exist. 
Let $\delta n$ denote the increment in the number of particles in an automaton-pointed segment 
when $\rho$ is increased by $\delta \rho$ from $\rho_{\rm th}$.
Then, we can estimate $\delta n= 4\delta\rho$ approximately 
where $4$ is the number of sites in one automaton-pointed segment.
Indeed, if $f$ is increased by $\delta f$ from $f=f_{\rm c}(\rho_{\rm th})$ satisfying
$\delta f>V_0(n+\delta n)(n+\delta n-1) - n(n-1)) = V_0(n\delta n+\delta n^2)$, 
where $n=1$ is the maximum particle number per one site at $f<f_{\rm c}(\rho_{\rm th})$, 
a particle can pass through an automaton-pointed segment in the driven direction even for the worst case. 
Thus, we obtain $f_{\rm c}(\rho)\simeq f_{\rm MF}(\rho)$ where 
\begin{eqnarray}
f_{\rm MF}(\rho)\equiv 4V_0(\rho-\rho_{\rm p}^{\rm l})+16V_0(\rho-\rho_{\rm p}^{\rm l})^2+f_{\rm c}(\rho_{\rm th}),
\end{eqnarray} for $\rho>\rho_{\rm p}^{l}$, otherwise $f_{\rm MF}(\rho)=0$.

Further, using (\ref{ff}), we can obtain the minimum possible value of
$f=f_0\equiv\frac{2\sqrt{2}V_0}{\sqrt{2}-1}\simeq 6.83$ 
to realize {\it MC}-global steady states at the dilute limit. 
At $f\ge f_0$, microscopic events change considerably,
 where a particle can overlap with the front particle in the driven direction 
without being affected by other particles. As a result, 
fluctuations in the local density could increased considerably, 
possibly causing another singularity in $\mathcal{J}$ even in finite density regions.
Thus, we can predict that there is a cross point 
$\rho_{0}\simeq 0.69$, determined by $f_{\rm MF}(\rho_{0})=f_0$, where the percolation effects 
and the large fluctuations of local densities become comparable.

\subsection{Conjecture B: Finite-size effects 
due to non-commutativity between $t\to\infty$ and $L\to\infty$}
Let us consider the situation with a sufficiently large driving field, 
where $N_{{\rm p},y}N_{{\rm n},y}\neq 0$ for some values of $y$ at a finite time $t$, 
and focus on such a line-$y$, which is a set of sites $\{i_x, y\}_{i_x}$. 
We consider the conditions where such all line-$y$
can be divided into {\it finite} number of segments, 
each of which satisfies $N_{{\rm p},y}^xN_{{\rm n},y}^x= 0$ 
where $N_{{\rm p}({\rm n}),y}^x$ is the number of positive 
(negative) particles in segment $x$ at $i_y=y$.
Clearly, there are no positive contributions to $J(t)$
at the contact points between such segments.
However, owing to the finite numbers of segments, 
we can neglect the effects brought about by such contact points with $L\to\infty$.
Hence, in the thermodynamic limit $L\to\infty$, ${\mathcal J}=1$ 
can be realized in other states besides global-{\it MC} steady states 
if we assume that the number of segments divided by $L$ goes to zero at any time.
Furthermore, when we consider the limit $L\to\infty$ at a fixed large value of $t$, 
such states appear not to relax to any global-{\it MC} steady states. 
This is because, in simple terms, the time required for 
the displacement of particles to form global-{\it MC} steady states would be at least 
of the order of $L$. Therefore, owing to the uniform configurations under the initial conditions, 
this type of transient state should be dominant for sufficiently large driving fields, 
as compared to the global-{\it MC} steady states, in the procedure where $t\to\infty$ after $L\to\infty$.

\section{Numerical simulations}\label{num}
Next, in order to verify the plausibility of the obtained results, 
we perform numerical simulations with $t=50000 \tau_0$ Monte Carlo steps 
and practically set $\tau=40000\tau_0$. Further, we set $V_0=1$ without loss of generality. 
We have verified that the following results are qualitatively 
unchanged with $\tau=20000\tau_0$. As shown in figure \ref{image1}, 
we have observed a global-{\it MC} steady state at $\rho=0.5$, $f=100.0$, and $L=50$.
It should be noted that observing global-{\it MC} steady states becomes 
considerably difficult for larger system sizes, which is discussed later.
On the other hand, as shown in figure \ref{image2}, 
although we have observed a stuck steady state at $\rho=0.5$, $f=1.0$, and $L=50$, 
the stuck steady state is not identical to the {\it CA}-stuck steady states.
Generally, some parts of {\it CA}-stuck steady states 
and other complicated configurations exist in the observed stuck steady state.
Nevertheless, maximum-current steady states and 
stuck steady states exist in different parameter regions, 
which indicates the existence of phase transitions.

\begin{figure}
\centering
\includegraphics[width=5.5cm,clip]{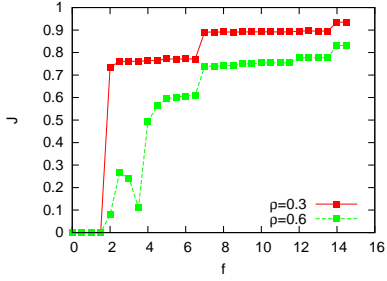}
\caption{Colored current ${\mathcal J}$ averaged in the time period 
from $\tau$ to $1.25\tau$ with $L=200$. 
There is a strong sample-to-sample fluctuation from $2< f < 4$ with $\rho=0.6$.}
\label{flow}
\end{figure}
\begin{figure}
\centering
\includegraphics[width=5.5cm,clip]{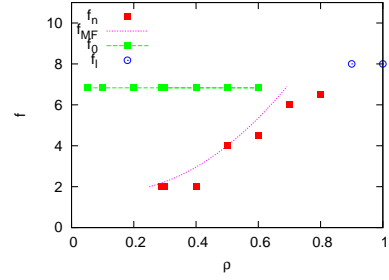}
\caption{Phase diagram estimated by $f_{\rm n}(\rho)$ numerically (points) 
and by $f_{\rm MF}(\rho)$ using the mean-field type analysis (real line). 
Observed weak singularities around $f_0$ in $\mathcal{J}$ and $f_{\rm l}$ are 
plotted (See the text for the definitions). The interval for fixed values of $\rho$ is $0.1$.
Further, the data with $\rho=0.29$ and $\rho=0.05$ are plotted 
because$f_n$ for $\rho\le 0.28$ and the singularity around $f_0$ for $\rho\le 0.04$ 
are not observed.}
\label{phase}
\end{figure}

In order to investigate the expected phase transitions explicitly, 
we change the external field by an increment of $\delta f = 0.5$ 
at a fixed density $\rho$ and measure 
$\Delta{\mathcal J}(f)\equiv \mathcal J(f)-\mathcal J(f-\delta f)$.
Then, we attempt to detect a point
\begin{eqnarray}
f_{\rm n}(\rho)\equiv {\rm arg~max}_{f} \Delta{\mathcal J}(f),
\end{eqnarray} which can be a reasonable candidate for the singular points 
$f_{\rm c}(\rho)$ of $\mathcal{J}$. 
It should be noted that the following observation is made 
by one simulation run under an initial condition 
with $L=200$. As shown in figure \ref{flow}, 
the numerical simulations demonstrate that it is easy to determine $f_{\rm n}(\rho)$ 
at sufficiently low densities because $\mathcal{J}$ jumps discontinuously 
at some values of $f$. 
As shown in the figure \ref{phase}, 
using $f_{\rm n}(\rho)$, we can approximately estimate $f_{\rm c}(\rho)$, 
which operationally defines a flowing phase and a blocked phase. 
It should be noted that
 another weak jump around $f_0$ was observed independent of $\rho$ 
at sufficiently low densities, as expected by the {\it Conjecture A}.
However, it is not easy to estimate the location of $f_{\rm n}(\rho)$ for $\rho\ge 0.5$ 
because of sample-to-sample fluctuations in $f_{\rm n}(\rho)$. For this region, we 
need another quantity for the plausible estimation of $f_{\rm n}(\rho)$, as discussed later.
Further, discontinuous jumps of $\mathcal{J}$ seem to disappear in high density regions 
such as $\rho=0.9$ and $1.0$. 

\begin{figure}
\centering
\includegraphics[width=5.5cm,clip]{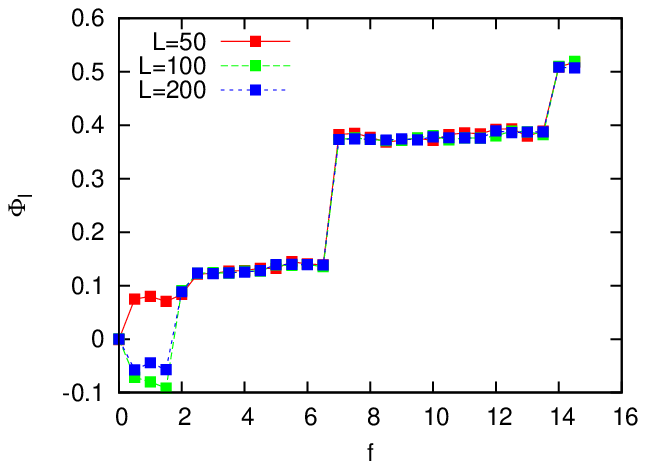}
\caption{The local lane order parameter $\Phi_{\rm l}$ averaged in the time period 
from $\tau$ to $1.25\tau$ for different system sizes at $\rho=0.3$.}
\label{local}
\end{figure}

\begin{figure}
\centering
\includegraphics[width=5.5cm,clip]{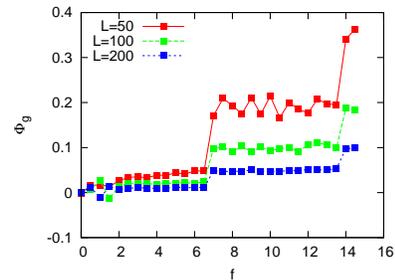}
\caption{The global lane order parameter $\Phi_{\rm g}$ averaged in the time period 
from $\tau$ to $1.25\tau$ for different system sizes at $\rho=0.3$.}
\label{global}
\end{figure}

Next, we investigate finite size effects in the system, in particular, 
focusing on lane formations.
Lane formations indicate that 
particles tend to be aligned to the direction of the driving field. 
The lane formations can be characterized by at least two approaches.
The first approach, referred to as local lane formation, 
involves estimating how often the same type of particles are nearest neighbors of the focused 
particle in the driven direction \cite{Lowen3}.
The second approach, referred to as global lane formation, 
involves estimating the number of times the same type of particles occur at all sites in the driven direction.
\cite{Lowen1,Lowen11,Lowen2}.
As one possible way to quantify such local lane formations, we focus on 
\begin{eqnarray}
\Phi_{\rm l}\equiv \phi_{\rm X}(l_o)-\phi_{\rm Y}(l_o),
\end{eqnarray}
where $\phi_{\rm X}(\l_0)\equiv \frac{1}{\rho L^2}\sum_{i\in \Lambda}|\sigma_i|
\frac{X_i^{+}-X_i^{-}}{X_i^{+}+X_i^{-}}$ and 
$\phi_{\rm Y}(l_0)\equiv \frac{1}{\rho L^2}\sum_{i\in \Lambda}|\sigma_i|
\frac{Y_i^{+}-Y_i^{-}}{Y_i^{+}+Y_i^{-}}$
with $X_i^{\pm} \equiv \sum_{i_x'=i_x-l_0}^{i_x+l_0}(|\sigma_{i_{x}',i_y}|-\delta_{i_{x}',i_x})
\Theta(\pm\sigma_{i_x,i_y}\sigma_{i_{x}',i_y})$ and
$Y_i^{\pm} \equiv \sum_{i_y'=i_y-l_0}^{i_y+l_0}(|\sigma_{i_{x},i_y'}|-\delta_{i_{y}',i_y})
\Theta(\pm\sigma_{i_x,i_y}\sigma_{i_{x},i_y'})$. 
Here, in order to determine nearest-neighbor particles for the focused particle, 
we set $l_0=\lfloor \rho^{-1} \rfloor$ for $L/2\ge \lfloor \rho^{-1} \rfloor$ 
under the assumption that there two particles exist in the segment at an average distance of 
approximately $2l_0$ in the driven direction. 
On the other hand, as one possible way to quantify global lane formations, we focus on 
\begin{eqnarray}
\Phi_{\rm g}\equiv\phi_{\rm X}(L/2).
\end{eqnarray} As shown in figure \ref{local}, 
$\Phi_{\rm l}$ shows a clear discontinuous jump between a negative value and a positive value 
around the obtained $f_{\rm n}(\rho)$ even at $\rho\ge 0.5$.
Further, another singularity around $f_0$ is clearer than that of ${\mathcal J}$.
In high-density regions such as $\rho=0.9$ or $1.0$, 
it is also clear that there is one large discontinuous jump between a negative value and a positive value
whose critical fields $f_{\rm l}$ are plotted in the figure \ref{phase} as a reference.
More importantly, the global tendencies of $\Phi_{\rm l}$ do not depend on the system sizes, 
which indicates that behaviors of local lane formations are independent of the size of the system. 
We have numerically observed that global tendency of $\mathcal{J}$ also do not depend 
on the system sizes at $f\ge f_{\rm l}(\rho)$, which is reasonable because $\mathcal{J}$ is also a local quantity.
Therefore, we used $f_{\rm l}(\rho)$ as the value of $f_{\rm n}(\rho)$ 
for $0.5\le\rho\le 0.8$ in the thermodynamic limit.
On the other hand, as shown in figure \ref{global}, 
$\Phi_{\rm g}$ shows a strong dependence on the system sizes. 
In particular, the tendency of these finite-size effects 
indicates that global lane formations disappear in the procedure 
where $t\to\infty$ after $L\to\infty$, which is consistent with {\it Conjecture B}. 

As shown in figure \ref{phase}, $f_{\rm MF}(\rho)$ and $\rho_0$
obtained by the mean-field type analysis in {\it Conjecture A} 
provide qualitatively consistent behaviours with $f_{\rm n}(\rho)$,
although the mean-field type analysis appears to overestimate the parameter regions of the blocked phase.
These deviations are rather natural for this mean-field type analysis 
because of the existence of spatial correlations that we have ignored.

\section{Concluding remarks}
We have proposed a simple lattice model for oppositely driven binary particles 
with purely repulsive interactions.
We have exactly constructed global-{\it MC} steady states 
at sufficiently large values of the driving field for any density 
and {\it CA}-stuck steady states at small driving fields for low densities. 
A mean-field type analysis leads us to estimate singular points in the colored current, 
which are qualitatively consistent with numerical simulations.
This strongly suggests that such a singularity originates from 
percolation in a stochastic cellular automaton buried in the equilibrium configuration.
Further, we have presented a conjecture for the absence of the relaxation toward
global-{\it MC} steady states (global lane formations) from the equilibrium initial conditions 
in the procedure where $t\to\infty$ after $L\to\infty$, 
which is also consistent with numerical simulations. 

Here, we present some remarks on the universality of the observed phenomena in the proposed model. 
First, we describe the singular point $f_0$ in low-density regions, 
above which the driving field is large enough to allow a particle to enter 
inside a soft-core repulsive potential via a two-body collision.
After such a strong collision, the fluctuations in the local density 
could be considerably strong. This type of strong density fluctuations 
might be related to instabilities of homogeneous density profiles
in a Langevin system \cite{Lowen2} and a two-lane traffic model \cite{Gregory}.
On the other hand, the Langevin system exhibits a lane formation 
also in high-density regions \cite{Lowen2}, 
which might be related to the singularities above $\rho_0$ in the proposed model.
Further, in the case of adding a finite-range repulsive interaction 
$\sum_{<ij>}|V_1\sigma_i\sigma_j|$ to the Hamiltonian, 
we have numerically found that the lane formation is qualitatively similar 
to those in this study if $|V_1|\ll 1$. 
Such extensions including finite temperatures are interesting 
when comparing the proposed model to the other more realistic systems 
with various effects such as inertial effects \cite{Del}, 
attractive interactions \cite{Lowen5}, hydrodynamic interactions \cite{Lowen6}.
Thus, the robustness of the universality of the observed singularities in the proposed model 
are interesting topics for future studies.

\acknowledgments 
The author thanks C. P. Royall for the discussions on papers \cite{Lek,Lowen3}
and also M. Ikeda, H. Wada, and H. Hayakawa for the collaboration at the initial stage of this work
when the author was a postdoctoral researcher in Yukawa Institute for Theoretical Physics (YITP). 
The author also thanks T. Matsumoto, M. Yamada, and G. Szamel for useful comments at the author's seminar at YITP.


\begin{thebibliography}{99}
\bibitem{Lowen1}
  \Name{Dzubiella J., Hoffman G. P., \and L\"owen H.}
  \REVIEW{Phys. Rev. E}{65}{2002}{021402}. 
\bibitem{Lowen11} 
\Name{Chakrabarti J., Dzuniella J., \and L\"owen H.}
\REVIEW{Europhys. Lett.}{61}{2003}{415}.
\bibitem{Lowen2} 
\Name{Chakrabarti J., Dzuniella J., \and L\"owen H.}
\REVIEW{Phys. Rev E}{70}{2004}{012401}.
\bibitem{Netz}
  \Name{Netz R. R.} 
\REVIEW{Europhys. Lett.}{63}{2003}{616}.
\bibitem{Lek} 
\Name{Leunissen	M. E. {\it et al.}}
\REVIEW{Nature}{437}{2005}{235}.
\bibitem{Lowen4}
\Name{S\"utterlin K. R. {\it et al.}}
\REVIEW{Phys. Rev. Lett.}{102}{2009}{085003}.
\bibitem{Lowen3}
\Name{Vissers T. {\it et al.}}
\REVIEW{Soft Matter}{7}{2011}{2352}.
\bibitem{Spohn} 
\Name{Spohn H.} 
\Book{Large scale dynamics of interacting particles} 
\Year{1991} \Publ{Springer-Verlag}.
\bibitem{Derrida}
\Name{Derrida B.}
\REVIEW{Phys. Rep.}{301}{1998}{65}.
\bibitem{Sasamoto}
\Name{Sasamoto T. \and Spohn H.}
\REVIEW{Phys. Rev. Lett.}{104}{2010}{230602}.
\bibitem{Hucht}
\Name{Hucht A.}
\REVIEW{Phys. Rev. E}{80}{2009}{061138}.
\bibitem{Hin}
\Name{Hinrichsen H.}
\REVIEW{Adv. Phys.}{49}{2000}{815}.
\bibitem{Gregory}
\Name{Appert-Rolland C., Hilhorst H. J, \and Schehr G.}
\REVIEW{J. Stat. Mech.}{}{2010}{P08024}.
\bibitem{Del}
\Name{Delhommelle J.}
\REVIEW{Phys. Rev. E}{71}{2005}{016705}.
\bibitem{Lowen5}
\Name{Rex M., \and L\"owen H.}
\REVIEW{Phys. Rev. E}{75}{2007}{051402}.
\bibitem{Lowen6}
\Name{Rex M., \and L\"owen H.}
\REVIEW{Eur. Phys. J. E}{26}{2008}{143}.
\end{thebibliography}
\end{document}